\documentclass[onecolumn,showpacs,preprintnumbers,nofootinbib,amsmath,amssymb]{revtex4}

\usepackage{amsfonts} 
\usepackage{amssymb}  
\usepackage{graphicx} 
\usepackage{amsmath}  
\usepackage{amsmath}  
\usepackage[latin1]{inputenc}
\usepackage{color}

\begin{document}
\title{Thermodynamical properties of a neutral vector boson gas in a constant magnetic field}
\author{G. Quintero Angulo\footnote{gquintero@fisica.uh.cu}}
\affiliation{Facultad de F{\'i}sica, Universidad de la Habana,\\ San L{\'a}zaro y L, Vedado, La Habana 10400, Cuba}
\author{A. P\'erez Mart\'{\i}nez\footnote{aurora@icimaf.cu}}
\affiliation{Instituto de Cibern\'{e}tica, Matem\'{a}tica y F\'{\i}sica (ICIMAF), \\
 Calle E esq a 15 Vedado 10400 La Habana Cuba\\}
\author{H. P\'erez Rojas\footnote{hugo@icimaf.cu}}
\affiliation {Instituto de Cibern\'{e}tica, Matem\'{a}tica y F\'{\i}sica (ICIMAF), \\
 Calle E esq a 15 Vedado 10400 La Habana Cuba\\}

\thanks{}%
\date{\today}%

\begin{abstract}
  The thermodynamical properties of a neutral vector boson gas in a constant magnetic field are studied starting from the spectrum given by Proca formalism. Bose Einstein Condensation (BEC) and  magnetization are obtained, for the three and one dimensional cases, in the limit of low temperatures.  In three dimensions the gas undergoes a phase transition to an usual BEC in which the critical temperature depends on the magnetic field. In one dimension a diffuse condensate  appears as for the charged vector boson gas. In both cases, the condensation is reached not only by decreasing the temperature but also by increasing the magnetic field. In three and one dimensions self-magnetization is possible. The anisotropy in the pressures due to axial symmetry imposed to the system by the magnetic field is also discussed. The astrophysical implications are commented.

\end{abstract}


\maketitle

\section{Introduction}	

There is a diversity of structures associated with a wide range of magnetic fields ($10^{-9}-10^{15}$~G) cohabiting in our Universe. Salient examples are galaxies (radius $1.5 \times 10^{18}~$km) and compact objects (radius $20~$km). The internal composition of neutron stars (still poorly understood) is described by all sorts of exotic dense matter in form of hyperons, a Bose Einstein Condensate (BEC) of mesons or deconfined quark matter in presence of strong magnetic fields \cite{Lattimer:2004pg}. Size and shape of a compact object depends on its composition but also on the magnetic field \cite{0954-3899-36-7-075202}. There are also some phenomena at astrophysical scale without explanation, as jets of pulsars, where magnetic fields might be relevant \cite{MNL2:MNL2848,Charbonneau:2009ax}.

Even though some theories have been proposed to explain the origin of such magnetic fields, this issue is far from being exhausted and it is still under great debate. In this regard, spin one bosons seems to be good candidates as magnetic field sources since, at low temperature, they are known to show an spontaneous magnetization. As a consequence, under certain conditions, a gas of bosons can generate and sustain its own magnetic field \cite{Yamada, Elizabeth}.

The study of BEC and magnetization for a charged scalar or vector boson gas in presence of a constant magnetic field was tackled in \cite{ROJAS1996148,PEREZROJAS2000,Khalilov:1999xd,Khalilov1997,Bargueno,Jian}. For low temperatures, the charged vector boson gas is paramagnetic, can be self magnetized and undergoes a phase transition to a diffuse BEC \cite{ROJAS1996148,PEREZROJAS2000}.

For a diffuse phase transition there is not a critical temperature, but an interval of temperatures along which it occurs gradually \cite{ROJAS1997}. In particular, a diffuse BEC phase is characterized by the presence of a finite fraction of the total particle density in the ground state and in states on its neighborhood at some temperature $T>0$. In this sense, the criterion for defining a diffuse BEC is weaker than the used for the usual one, which requires  the existence of a critical temperature below which there is a macroscopic amount of particles in the ground state. Whether a BEC is usual or diffuse is strongly related to the dimension of the system \cite{ROJAS1997}.


Although both, charged and neutral vector bosons, could be relevant participants of astronomical phenomena, the thermodynamics of the neutral vector boson gas has been less studied. An effect analogous to self-magnetization, named BE-ferromagnetism, was founded in \cite{Yamada} for a gas of non relativistic ideal neutral boson with spin one. In \cite{Chernodub:2010qx} and \cite{Chernodub:2012fi} magnetic field induced superconductivity and superfluidity are obtained for a gas of charged an neutral vector mesons, but it ignores the weak coupling between the neutral mesons and the magnetic field. More recently, BEC for an gas of interacting vector bosons at zero magnetic field was studied in \cite{Satarov:2017jtu}.

Hence, the aim of this paper is to study the thermodynamical properties of a neutral vector boson  gas (NVBG) in a constant magnetic field. We will concern with its phenomenology in the framework of Proca theory and independently of the realistic conditions in which it may appear. Neutral vector bosons can be mesons, atoms and other paired fermions with total integer spin. For numerical calculations we use a positronium gas parameters, characterized by a mass approximately $2 m_e$, ($m_e$ is the electron mass) and twice the electron magnetic moment $\kappa = 2\mu_B $ being $\mu_B$ the Bohr magneton. Since we are focussed on possible astrophysical applications we deal with systems of densities in the range of $10^{30}-10^{34}cm^{-3}$.

The thermodynamical properties of the NVBG are studied in three and one dimensions. The BEC phase transition turns out to be diffuse in one dimension and usual in three. In both cases, for low enough temperatures self-magnetization arises. An analysis of this phenomenon lead us to the conditions for the appearance of a self sustained field. The axial symmetry of the magnetic field is reflected in the particle spectra and in the energy-momentum tensor of the system which becomes anisotropic. For that reason, we also study the splitting of the parallel and perpendicular pressures with respect to the direction of the magnetic field.

Our paper is organized as follows. In Section~\ref{eom}, we present the equation of motion and spectrum of neutral vector boson with MM. Section~\ref{thermo} contains a derivation of the thermodynamical potential, particle density,  BEC, internal energy, entropy and specific heat for the three and one dimensional  NVBG. Magnetization, self-magnetization and anisotropic pressures are also discussed.  Section~\ref{conclusions} is devoted to conclusions. Appendix A and B contains some details of the calculations.

\section{Equation of motion of a neutral vector boson bearing a magnetic moment}
\label{eom}

 Neutral spin-one bosons with magnetic moment that moves in a magnetic field can be described by an extension  of the original Proca  Lagrangian for spin one particles that includes particle-field interactions \cite{PhysRev.131.2326,PhysRevD.89.121701}

\begin{eqnarray}\label{Lagrangian}
  L = -\frac{1}{4}F_{\mu\nu}F^{\mu\nu}-\frac{1}{2} \rho^{\mu\nu}\rho_{\mu\nu}
       + m^2 \rho^{\mu}\rho_{\mu}
      +i m \kappa(\rho^{\mu} \rho_{\nu}-\rho^{\nu}\rho_{\mu}) F_{\mu\nu}.
\end{eqnarray}

In Eq.~(\ref{Lagrangian}) the indices $\mu$ and $\nu$ run from 1 to 4, $F^{\mu\nu}$ is the electromagnetic tensor, and  $\rho_{\mu\nu}$, $\rho_{\mu}$ are independent field variables that follow \cite{PhysRev.131.2326}

\begin{equation}\label{fieldeqns}
  \partial_{\mu} \rho_{\mu\nu}-m^2 \rho_{\nu}+ 2i \kappa m \rho_{\mu} F_{\mu\nu}=0,\quad\quad
  \rho_{\mu\nu} = \partial_{\mu} \rho_{\nu} - \partial_{\nu} \rho_{\mu}.
\end{equation}

A variation of the Lagrangian  with respect to the field $\rho_{\mu}$ give us the equations of motion that in the momentum space read


%
%
%

\begin{equation}
\left((p_{\mu}^2  + m^2)\delta_{\mu\nu} -p_{\mu} p_{\nu}  - 2  i \kappa m F_{\mu \nu}\right)\rho_{\mu} = 0.
 \end{equation}

Thus, the boson propagator is

\begin{equation}\label{propagator}
D_{\mu\nu}^{-1}=((p_{\mu}^2  + m^2)\delta_{\mu\nu}-p_{\mu} p_{\nu}  - 2  i \kappa m F_{\mu \nu}).
\end{equation}

Considering the magnetic field  uniform, constant and in $p_3$ direction $\textbf{B}=B\textbf{e}_3$ one can start from Eq.~(\ref{fieldeqns}) and obtain the generalized Sakata-Taketani hamiltonian for the six component wave equation of the system \cite{PhysRev.131.2326, PhysRevD.89.121701} following the same procedure of Ref. \cite{PhysRev.131.2326}. The hamiltonian reads

\begin{equation}\label{hamiltonian}
H = \sigma_3 m + (\sigma_3 + i \sigma_2) \frac{\textbf{p}^2}{2 m} -
    i \sigma_2 \frac{(\textbf{p}\cdot\textbf{S})^2}{m}
    -(\sigma_3 - i \sigma_2) \kappa \textbf{S} \cdot \textbf{B},
\end{equation}

\noindent with $\textbf{p}=(p_{\perp},p_3)$ and $p_{\perp}=p_1^2 + p_{2}^2$.  $\sigma_{i}$ are the $2\times2$ Pauli
matrices, $S_{i}$ are the $3\times3$ spin-1 matrices in a representation in which $S_3$ is diagonal and $\textbf{S} = \{S_1,S_2,S_3\}$\footnote{
$\begin{array}{ccc} S_1=\frac{1}{\sqrt{2}}
\left( \begin{array}{ccc}
0 & 1& 0\\
1 & 0 & 1\\
0 & 1 & 0
\end{array}\right),
& S_2=\frac{i}{\sqrt{2}}
\left( \begin{array}{ccc}
0 & \text{-}1& 0\\
1 & 0 & \text{-}1\\
0 & 1 & 0
\end{array} \right),
& S_3=
\left( \begin{array}{ccc}
1 & 0& 0\\
0 & 0 & 0\\
0 & 0 & \text{-}1\end{array}\right)\end{array}$}.


%
%
%
%
%


The eigenvalues of (\ref{hamiltonian}) are
\begin{equation}
\varepsilon(p_3,p_{\perp}, B,s)=\sqrt{m^2+p_3^2+p_{\perp}^2-2\kappa s B\sqrt{p_{\perp}^2+m^2}},\label{spectrum}
\end{equation}

\noindent where $s=0, \pm 1$ are the spin eigenvalues. Although we are dealing with neutral bosons, as happens for the charged ones, the magnetic field intensity $B$ enters in the energy coupled with the perpendicular momentum (see the last term in the previous equation). This coupling reflects the axial symmetry imposed to the system by the magnetic field.

The ground state energy of the neutral spin one boson ($s=1$ and $p_3=p_{\perp}=0$) is
\begin{equation}
\varepsilon(0, B)=\sqrt{m^2-2\kappa B m}=m\sqrt{1-b}.\label{massrest}
\end{equation}
\noindent  being $b=\frac{B}{B_c}$ and $B_c=\frac{m}{2\kappa}$.

From Eq.~(\ref{massrest}) follows that the rest energy of the system decreases with the magnetic field and becomes zero for $B=B_c$. For the values or $m$ and $\kappa$ we are considering, $m=2m_e$ and $\kappa=2\mu_B$, $B_c=m_e^2/e=4.41 \times 10^{13}~G$ which is the Schwinger critical field. Let us note that the ground state of the charged vector boson has a similar instability (see \cite{ROJAS1996148}).


\noindent Eq.(\ref{massrest}) allows us to define and effective magnetic moment as

\begin{equation}
d=-\frac{\partial \varepsilon(0,B)}{\partial B}=\frac{\kappa m}{\sqrt{m^2-2m\kappa B}}=\frac{\kappa}{\sqrt{1-b}}\label{magmoment}.
\end{equation}

The system has a paramagnetic behavior because $d>0$. It will be also important for the discussion below the fact that $d$ grows with the increasing of the magnetic field and diverges for $b \rightarrow 1$ $(B \rightarrow B_c)$.

\section{Thermodynamical properties}
\label{thermo}
The general expression for the thermodynamical potential of the NVBG has the form

 \begin{equation}\label{Grand-Potential-4}
\Omega(B,\mu,T)= \sum_{s=-1,0,1}\frac{1}{\beta}\left[\sum_{p_4}\int\limits_{-\infty}^{\infty}\frac{p_{\perp}dp_{\perp}dp_3}{(2\pi)^2} \ln \det D^{-1}(\overline{p}^*)\right].
\end{equation}
Here $D^{-1}(\overline{p}^*)$ is the neutral boson  propagator given by  (\ref{propagator}),  $\beta = 1/T$  denotes the inverse temperature, $\mu$ the boson chemical potential and ${\overline{p}}^*=(ip_{4}-\mu,0,p_{\perp},p_{3})$.
After doing the Matsubara sum, Eq.(\ref{Grand-Potential-4}) becomes

\begin{equation}
\Omega(B,\mu,T)= \Omega_{st}+\Omega_{vac},
\end{equation}

\noindent where $\Omega_{st}$ is the statistical contribution of bosons/antibosons that depends on $B, T$ and $\mu$

\begin{equation}
\Omega_{st}(B,\mu,T)=  \sum_{s=-1,0,1} \frac{1}{\beta}\left(\int\limits_{0}^{\infty}\frac{p_{\perp}dp_{\perp}dp_3}{(2\pi)^2} \ln \left((1-e^{-(\varepsilon(p_3,p_{\perp}, B,s)- \mu)\beta})(1-e^{-(\varepsilon(p_3,p_{\perp}, B,s)+ \mu)\beta})\right)  \right ),
\end{equation}

\noindent and $\Omega_{vac}$  is the vacuum term which is only B-dependent \footnote{The vacuum term is important for instance, in the positronium case. It would represent a correction to the usual Euler-Heisenberg term in which the electron positron pairs bosonize by coupling, for instance, through Coulomb force.}

\begin{equation}
\Omega_{vac}=\sum_{s=-1,0,1}\int\limits_{0}^{\infty}\frac{p_{\perp}dp_{\perp}dp_3}{(2\pi)^2}\varepsilon(p_3,p_{\perp}, B,s).
\end{equation}

We can rewrite $\Omega_{st}$ as

\begin{equation}
\Omega_{st}(B,\mu,T)=  \sum_{s=-1,0,1} \Omega_{st}(s),
\end{equation}

\noindent with

\begin{equation}\label{Grand-Potential-sst}
\Omega_{st}(s)=\frac{1}{\beta}\left(\int\limits_{0}^{\infty}\frac{p_{\perp}dp_{\perp}dp_3}{(2\pi)^2} \ln \left((1-e^{-(\varepsilon(p_3,p_{\perp}, B,s)- \mu)\beta})(1-e^{-(\varepsilon(p_3,p_{\perp}, B,s)+ \mu)\beta})\right)  \right )
\end{equation}

\noindent being the contribution of each spin state to the statistical part of the potential.


Using the Taylor expansion of the logarithm, Eq.~(\ref{Grand-Potential-sst}) can be written as

\begin{equation}\label{Grand-Potential-sst1}
\Omega_{st}(s)= - \frac{1}{4 \pi^2 \beta}  \sum_{n=1}^{\infty} \frac{e^{n \mu \beta}+e^{- n \mu \beta }}{n} \int\limits_{0}^{\infty} p_{\perp}  dp_{\perp} \int\limits_{-\infty}^{\infty} dp_3 e^{-n \beta \varepsilon(p_3,p_{\perp}, B,s)}.
\end{equation}

\noindent The term $e^{n \mu \beta}$ stands for the particles and the term $e^{- n \mu \beta}$ for the antiparticles.

\noindent In Eq.~(\ref{Grand-Potential-sst1}) the integration in $p_3$ can be completely carried out, while the integration in $p_{\perp}$ can be only partially done and we obtain for the thermodynamical potential the expression

\begin{eqnarray}\label{Grand-Potential-sst2}
\Omega_{st}(s)= - \frac{z_0^2}{2 \pi^2 \beta^2}  \sum_{n=1}^{\infty} \frac{e^{n \mu \beta}+e^{- n \mu \beta }}{n^2} K_2 (y z_0)
- \frac{\alpha}{2 \pi^2 \beta}  \sum_{n=1}^{\infty} \frac{e^{n \mu \beta}+e^{- n \mu \beta }}{n} \int\limits_{z_0}^{\infty} dz
\frac{z^2}{\sqrt{z^2+\alpha^2}} K_1 (y z),
\end{eqnarray}

\noindent where $K_n(x)$ is the McDonald function of order $n$, $ y = n \beta$, $z_0= m \sqrt{1-s b}$ and $\alpha=s m b/2$.

\noindent To compute the integral in the second term of Eq.~(\ref{Grand-Potential-sst2})

\begin{eqnarray}
I = \int\limits_{z_0}^{\infty} dz
\frac{z^2}{\sqrt{z^2+\alpha^2}} K_1 (y z),
\end{eqnarray}

\noindent we follow the procedure described in Appendix A. Finally   $\Omega_{st} (s)$ reads

\begin{eqnarray}\label{Grand-Potential-sst3}
\Omega_{st}(s)= - \frac{z_0^2}{2 \pi^2 \beta^2} (1+\frac{\alpha}{\sqrt{z_0^2 + \alpha^2}}) \sum_{n=1}^{\infty} \frac{e^{n \mu \beta}+e^{- n \mu \beta }}{n^2} K_2 (y z_0)
- \frac{\alpha z_0^2}{ \pi^2 \beta^2 \sqrt{z_{0}^2 + \alpha^2}} \\ \times \sum_{n=1}^{\infty} \frac{e^{n \mu \beta}+e^{- n \mu \beta }}{n^2} \sum_{w=1}^{\infty} \frac{(-1)^w(2 w -1)!!}{(z_{0}^2 + \alpha^2)^w} \left(\frac{z_0}{y}\right)^w K_{-(w+2)}(y z_0).\nonumber
\end{eqnarray}

In the low temperature limit $T \ll m$ (which for $m=2 m_e\cong1 MeV$ means $T \ll 10^{10} K$),  $\Omega_{st}(-1)$ and $\Omega_{st}(0)$ vanish. Therefore, in this limit  $\Omega_{st} \simeq \Omega_{st}(1)$. This means that all the particles are in the spin ground state $s=1$.
The leading term of $\Omega_{st}(1)$  is the first one in Eq.(\ref{Grand-Potential-sst3}).
Since it admits a further simplification, for the assumed low temperatures the statistical part of the thermodynamical potential is

\begin{eqnarray}\label{Grand-Potential-stfinal}
\Omega_{st}(B,\mu,T) = -\frac{\varepsilon(0,B)^{3/2}}{2^{1/2} \pi^{5/2} \beta^{5/2} (2-b)} Li_{5/2}(e^{\beta \mu^{\prime}}),
\end{eqnarray}

\noindent where  $Li_{n}(x)$ is the polylogarithmic function of order $n$ and $ \mu^{\prime} = \mu - \varepsilon(0,B)$. The quantity $\mu^{\prime}$ is a function of the temperature and the magnetic field, and it leads to the critical condition because the existence of a non-zero temperature $T_{cond}$ for which  $\mu^{\prime} = 0$ is the requirement for the usual BEC.

The vacuum contribution to the thermodynamical potential after being regularized is (see appendix B)
\begin{align}\label{Grand-Potential-vac}
\Omega_{vac} &= -\frac{m^4}{288 \pi}\left( b^2(66-5 b^2)-3(6-2b-b^2)(1-b)^2 \log(1-b)\right.
\\\nonumber
& \left. -3(6+2b-b^2)(1+b)^2\log(1+b) \right).
\end{align}

Adding Eqs.~(\ref{Grand-Potential-Tcero}) and (\ref{Grand-Potential-vac}) we get the total thermodynamical potential for the NVBG in the limit of low temperatures Eq.(\ref{Grand-Potential-Tcero})

\begin{align}\label{Grand-Potential-Tcero}
\Omega(B,\mu,T) & =-\frac{\varepsilon(0,B)^{3/2}}{2^{1/2} \pi^{5/2} \beta^{5/2} (2-b)} Li_{5/2}(e^{\beta \mu^{\prime}})-\frac{m^4}{288 \pi}\left( b^2(66-5 b^2) \right. \\\nonumber
& \left. -3(6-2b-b^2)(1-b)^2 \log(1-b)-3(6+2b-b^2)(1+b)^2\log(1+b) \right).
\end{align}

\subsection{Particle density and Bose Einstein Condensation}

To obtain the particle density we compute the derivative of Eq.(\ref{Grand-Potential-Tcero}) with respect to the chemical potential $\mu$

\begin{equation}\label{Particle-Density-Tcero}
N = -\frac{\partial \Omega(B,\mu,T)}{\partial \mu}= \frac{\varepsilon(0,B)^{3/2}}{2^{1/2} \pi^{5/2} \beta^{3/2} (2-b)} Li_{3/2}(e^{\mu^\prime  \beta}).
\end{equation}

This expression allows the substitution $\mu^{\prime} = 0$ (because $Li_{3/2}(1)=\zeta(3/2)$, where $\zeta(x)$ is the Riemann zeta function). Consequently, a critical temperature can be defined (Eq.~\ref{Tcond}) and the neutral vector  boson gas shows usual Bose Einstein Condensation

\begin{equation}\label{Tcond}
T_{cond} = \frac{1}{\varepsilon(0,B)} \left ( \frac{2^{1/2} \pi^{5/2} (2-b) N}{\zeta(3/2)}\right)^{2/3}.
\end{equation}

Although this behaviour resembles the one obtained for bosons at zero magnetic field -the functional relation between $T_{cond}$ and $N$ is the same-, when the field is present the critical temperature depends on it (through $\varepsilon(0,B)$ and $b$), and, what is more interesting, $T_{cond}$ diverges when $b \rightarrow 1$ ($B \rightarrow B_c$). The dependence of the critical temperature on the field put in evidence that the condensation can be reached not only by decreasing the temperature or augmenting the density, but also by increasing the magnetic field. This can be easily seeing if we compute the density of particles out of the condensate $N_{oc}$ (for $T<T_{cond}$)

\begin{equation}\label{Particle-Density-oc}
N_{oc} = \frac{\varepsilon(0,B)^{3/2}  T^{3/2}}{2^{1/2} \pi^{5/2} (2-b)} Li_{3/2}(e^{\mu^\prime  \beta}) = N \left(\frac{T}{T_{cond}}\right)^{3/2},
\end{equation}

\noindent because from Eq.~(\ref{Particle-Density-oc}) follows that $N_{oc} \rightarrow 0$ when $T \rightarrow 0$ but also when $b \rightarrow 1$
($ \varepsilon(0,B)\rightarrow 0$).

Given that a critical temperature is well defined for each value of the field (as well as there is a critical field for each temperature), it is possible to draw a $T$ vs $b$  phase diagram. We did so in Fig.~1 for two fixed values of the densities: $N=10^{30} cm^{-3}$ and $N=10^{32} cm^{-3}$,  respectively. These high boson densities  may be assumed for compact objects. The values of the critical temperatures (the dotted lines that separate the region where the BEC appears from that where there is no  BEC) are in the range of $T=10^7-10^9$~K which are also typical of compact objects. We can see from the graphics how $T_{cond}$ grows with the augment of the density and diverges when $b\rightarrow1$ ($B\rightarrow B_c$).

\begin{figure}[h]
\centering
\includegraphics[width=0.49\linewidth]{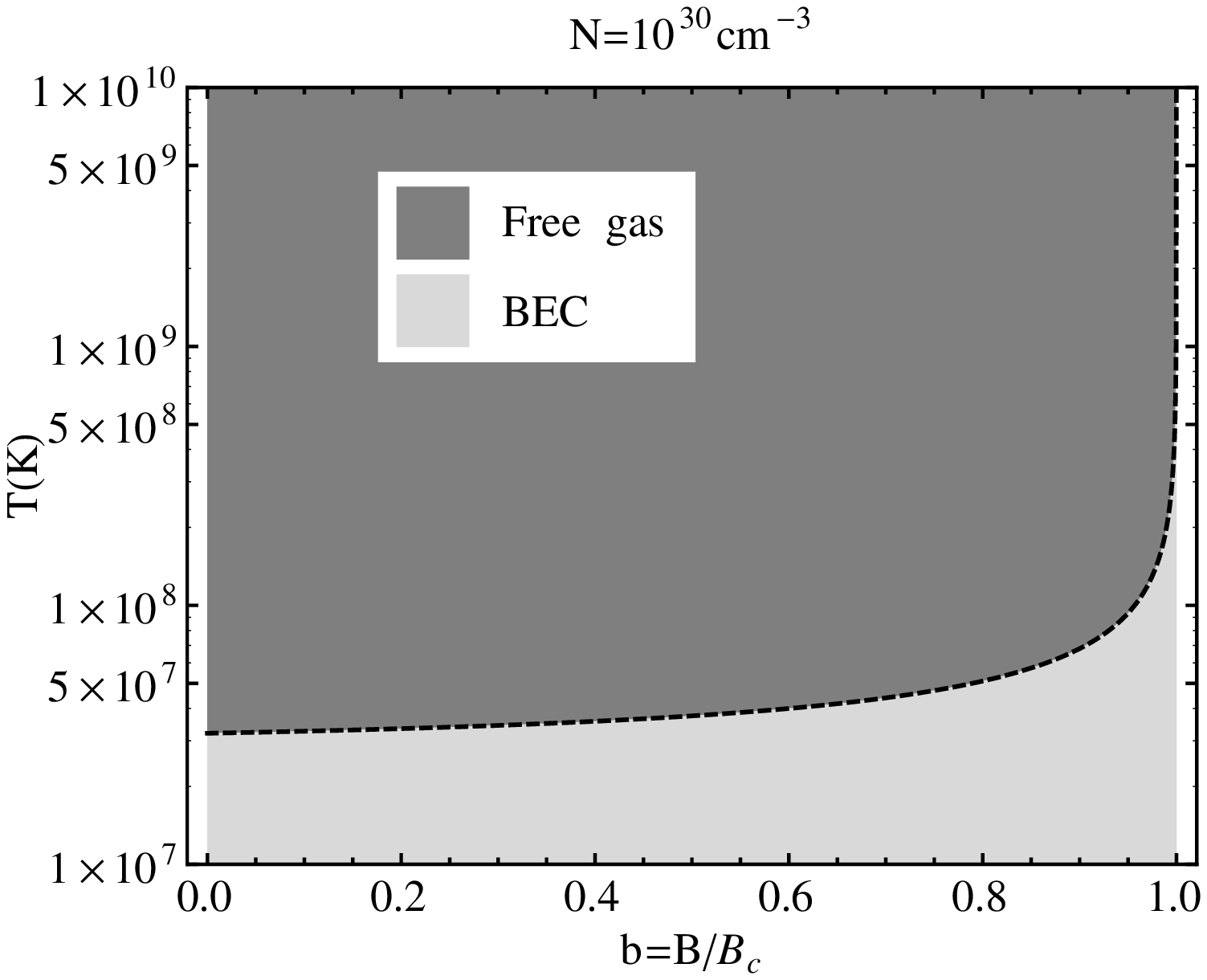}
\includegraphics[width=0.49\linewidth]{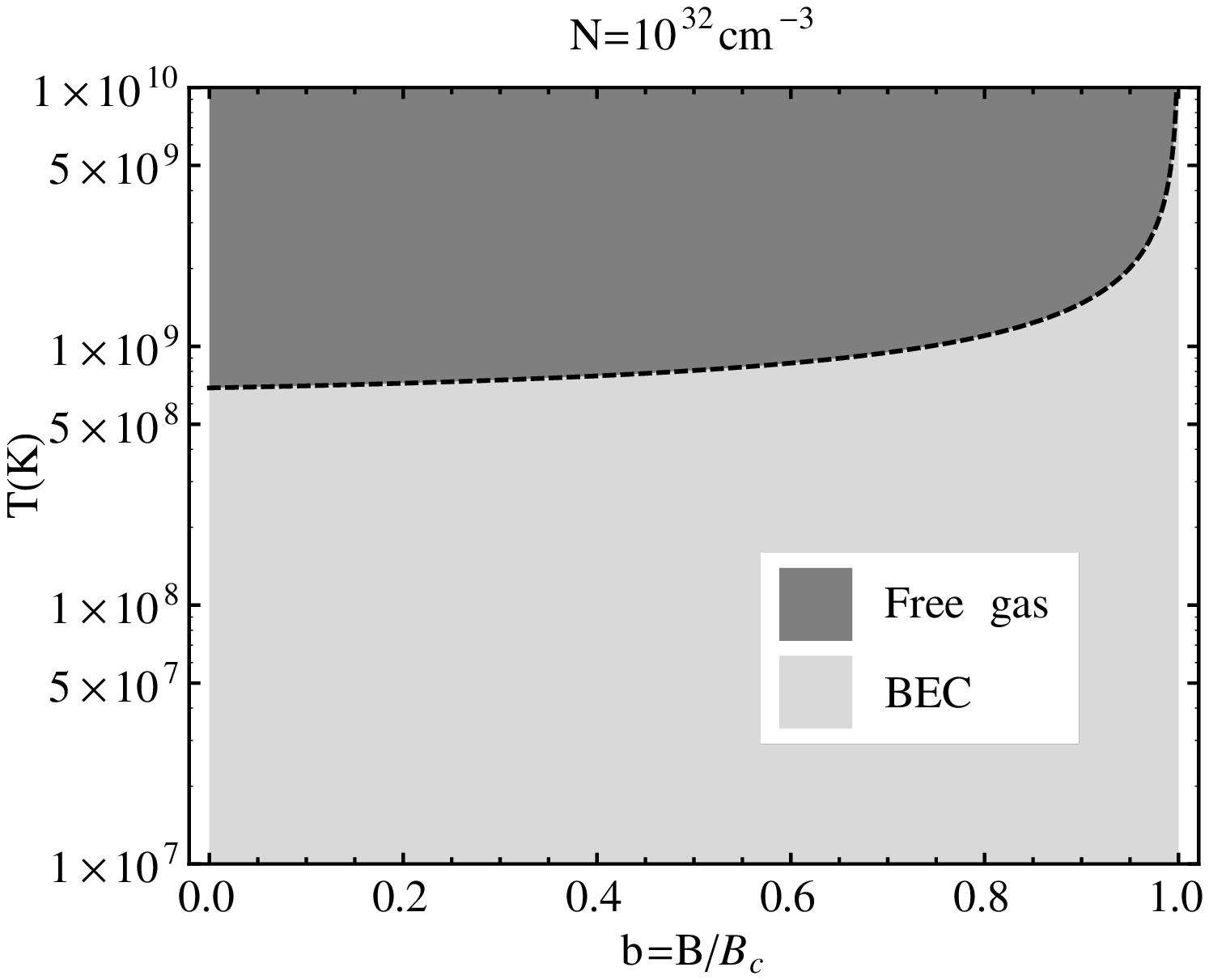}	
\caption{\label{fig5} The phase diagram in the $T-b$ space for different values of particle density. The black dashed lines are the critical curves that separates the region of $T$ and $b$ when there is condensate (light gray region) from the region in which there is not (dark gray region). Note that the critical line also depends on the particle density.}
\end{figure}

We can also examine the transition to the condensate through the behavior of specific heat. In particular we consider the specific heat at constant volume, defined by

\begin{equation}\label{specificheat}
C_v = \frac{\partial E}{\partial T},
\end{equation}

\noindent where $E = T S+\Omega+\mu N$ is the internal energy of the system.
The entropy of the vector bosons gas is

\begin{equation}\label{entropy}
S=-\frac{\partial \Omega}{\partial T}= -\beta \left(\mu^{\prime} N +\frac{5}{2} \Omega_{st} + \beta \frac{\partial \mu^{\prime}}{\partial \beta} N\right ),
\end{equation}

\noindent with

\begin{equation}
\mu^{\prime} \cong -\frac{\zeta(3/2)T}{4 \pi} \left ( 1-\left(\frac{T_{cond}}{T}\right)^{3/2} \right )\Theta(T-T_{cond})
\end{equation}

\noindent in the low temperature limit. Here $\Theta(x)$ is the Heaviside theta function.

With the use of Eq.~(\ref{entropy}) the internal energy can be written as

\begin{equation}\label{energy}
E = \varepsilon(0,B) N + \Omega_{vac}- \frac{3}{2} \Omega_{st} -\beta \frac{\partial \mu^{\prime}}{\partial \beta} N.
\end{equation}

Eq.~(\ref{energy}) allows us to obtain for the specific heat the following expression

\begin{equation}\label{specificheat-1D}
C_v = -\beta \left(\frac{15}{4} \Omega_{st} + \frac{3}{2} \mu^{\prime} N +\frac{1}{2} \beta \frac{\partial \mu^{\prime}}{\partial \beta} N- \beta^2 \frac{\partial^2 \mu^{\prime}}{\partial \beta^2}N\right ).
\end{equation}

The specific heat has been plotted in Fig.~2 as a function of the temperature for a fixed value of the density $N=10^{32} cm^{-3}$ and three values of the magnetic field. As it is apparent, it has a maximum that is a fingerprint of the BEC phase transition. The maximum decreases and smoothes with the increment of the magnetic field, and is expected to disappear for $B \rightarrow B_c$ ($b \rightarrow 1$), because when $B=B_c$ the gas is condensed at any temperature and density.

\begin{figure}[h]
\centering
\includegraphics[width=0.5\linewidth]{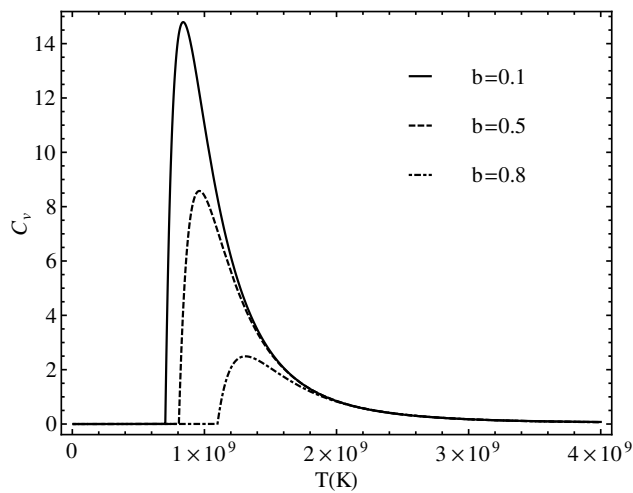}
\caption{\label{fig2} The specific heat as a function of the temperature for $N=10^{32} cm^{-3}$ and several values of the magnetic field.}
\end{figure}

At this point, it is worthwhile to comment on the reasons for the difference in the nature of the BEC showed by the charged and the neutral vector boson gas in a constant magnetic field. As it was already known, the BEC in the charged gas is diffuse \cite{ROJAS1996148,PEREZROJAS2000}, while we just found it is usual for the neutral gas. The difference arises from the reduction in the dimension suffered by the charged gas for lows temepratures or strong fields. This reduction is a consequence of the quantized spectrum, since the perpendicular  momentum component is replaced by Landau levels due to the coupling between the charge and the magnetic field. In the low temperature/high field limit the charged bosons concentrate in the lowest Landau Level (LLL) i.e, $n=0$ (which implies $p_{\perp}=0$), and therefore the three dimensional charged boson gas behaves as a one dimensional system \cite{ROJAS1996148}. Once the dimension is reduced, a further decrease in the temperature, or a increase in the magnetic field increases the population of the states around $p_3 \approx 0$ until it becomes a macroscopic quantity. But, due to the one dimensionality of the system, a critical temperature can not be defined, and the condensation is diffuse  \cite{ROJAS1997}.

For the neutral boson gas, the coupling between the field and the spin magnetic moment does not imply quantization for any momentum component. All of them are preserved as quantum observable and the three dimensionality of the neutral gas is kept for any temperature and magnetic field. For a Bose gas in three dimensions a critical temperature is well defined, and that is why it undergoes a phase transition to a usual BEC. However, it is still possible to find a formal analogy between the charged and the neutral spin-one gas in what respect to the condensation if we restrict the later to move in one dimension.

\subsubsection{Condensation in the one dimensional system}

Let us consider a  gas of neutral vector bosons with $p_{\perp}=0$. The statistical contribution to the thermodynamical potential has the form

%
%
%

\begin{equation}\label{Grand-Potential-2Dsst1}
\Omega^{1D}_{st}(B,\mu,T)= -\sum_{s=-1,0,1} \frac{1}{2 \pi \beta}  \sum_{n=1}^{\infty} \frac{e^{n \mu \beta}+e^{- n \mu \beta }}{n} \int\limits_{0}^{\infty} dp_3 e^{-n \beta \varepsilon(s)},
\end{equation}

\noindent where, as for the three dimensional gas, the terms $e^{n \mu \beta}$ and $e^{- n \mu \beta}$ stand for the particles and the antiparticles respectively. The integration over $p_3$ can be carried out, and we obtain for $\Omega^{1D}_{st}(B,\mu,T)$ the expression

\begin{equation}\label{Grand-Potential-2Dsst3}
\Omega^{1D}_{st}(B,\mu,T)= -\sum_{s=-1,0,1} \frac{m \sqrt{1- s b}}{ \pi \beta}  \sum_{n=1}^{\infty} \frac{e^{n \mu \beta}+e^{- n \mu \beta }}{n} K_1(n \beta m \sqrt{1-s b}).
\end{equation}

\noindent $K_1(x)$ is the first order McDonald function.

\noindent Performing the sum over the spin we have

\begin{eqnarray}\label{Grand-Potential-2Dstexact}
\Omega^{1D}_{st}(B,\mu,T)= - \frac{m \sqrt{1+ b}}{ \pi \beta}  \sum_{n=1}^{\infty} \frac{e^{n \mu \beta}+e^{- n \mu \beta }}{n} K_1(n \beta m \sqrt{1+b})\\ \nonumber
-\frac{m}{ \pi \beta}  \sum_{n=0}^{\infty} \frac{e^{n \mu \beta}+e^{- n \mu \beta }}{n} K_1(n \beta m)\\ \nonumber
-\frac{m \sqrt{1-b}}{ \pi \beta}  \sum_{n=0}^{\infty} \frac{e^{n \mu \beta}+e^{- n \mu \beta }}{n} K_1(n \beta m \sqrt{1-b}). \nonumber
\end{eqnarray}


\noindent As for the three dimensional gas, in the low temperature limit the leading term of Eq.~(\ref{Grand-Potential-2Dstexact}) comes from the particles with $s=1$. This leading term can be re-written as

\begin{equation}\label{Grand-Potential-2DstTcero}
\Omega^{1D}(B,\mu,T)= - \frac{\sqrt{\varepsilon(0,B)}}{\sqrt{2 \pi} \beta^{3/2}} Li_{3/2}(e^{\mu^{\prime 1D} \beta}),
\end{equation}

\noindent with $\mu^{\prime 1D} =\mu-\varepsilon(0,B)$. (Besides they have the same definition, to avoid a confusion, the distinction between $\mu^{\prime}$ for three dimensions and $\mu^{\prime 1D}$ is needed.)

The vacuum contribution to the 1D-thermodynamical potential after regularization is (see Appendix B)

%

\begin{equation}\label{Grand-Potential-2Dvacreg}
\Omega^{1D}_{vac}(B)= -\frac{m^2}{2 \pi} ((1-b) \log (1-b)+ (1+b)\log (1+b)),
\end{equation}

\noindent and finally for the 1D-thermodynamical potential in the limit of low temperature we have

\begin{equation}\label{Grand-Potential-2DTcero}
\Omega^{1D}(B,\mu,T) =  - \frac{\sqrt{\varepsilon(0,B)}}{\sqrt{2 \pi} \beta^{3/2}} Li_{3/2}(e^{\mu^{\prime 1D} \beta}) -\frac{m^2}{2 \pi} \left ( (1-b) \log (1-b)+ (1+b)\log (1+b) \right ).
\end{equation}

\noindent The particle density obtained from Eq.~(\ref{Grand-Potential-2DTcero}) is

\begin{equation}\label{Particle-Density-2DTcero}
N^{1D} = -\frac{\partial \Omega^{1D}(B,\mu,T)}{\partial \mu}= \frac{\sqrt{\varepsilon(0,B)}}{\sqrt{2 \pi \beta^{1/2}}} Li_{1/2}(e^{\mu^{\prime 1D}  \beta}).
\end{equation}

This expression is similar to the one obtained for a three dimensional charged vector boson gas in a constant magnetic field (see Eq.~(23) of  \cite{Khalilov1997}). It does not admit the substitution $\mu^{\prime 1D} = 0$ (because $Li_{1/2}(1)\rightarrow\infty$), therefore, a critical temperature can not be defined and the gas does not have BEC in the usual sense. Nevertheless, in the very low temperature limit, in which $\mu\sim \varepsilon(0,B)\gg T$ but $T \gg \mu^{\prime 1D}$ expression (\ref{Particle-Density-2DTcero}) can be approximated as

\begin{equation}\label{Particle-Density-2Dcondensate}
N^{1D} \simeq \frac{1}{2 \beta} \sqrt{\frac{2 \varepsilon(0,B)}{-\mu^{\prime 1D}}}.
\end{equation}

Again, an expression equivalent to  Eq.~(\ref{Particle-Density-2Dcondensate}) for a charged vector bosons gas has been already obtained in Ref.~\cite{ROJAS1996148}. Eq.~(\ref{Particle-Density-2Dcondensate}) has a divergence when $\mu'\rightarrow 0$ but the particle density must remain finite, thus its right interpretation is as a manner to compute $\mu^{\prime 1D}$

\begin{equation}\label{miuprima-1D}
\mu^{\prime 1D} = -\varepsilon(0,B)\frac{T^2}{ N^2}= -m \sqrt{1-b}\frac{T^2}{ N^2}.
\end{equation}

From Eq.~(\ref{miuprima-1D}) follows that $\mu^\prime$ is a decreasing function of T, as occurs in the usual Bose-Einstein condensation, but also that $\mu^{\prime 1D}$ is a decreasing function of the field that goes to zero when $B \rightarrow B_c$. Since $\mu^{\prime 1D} = 0$ is the condition for the BEC to occur, $\mu^{\prime 1D} \sim 0$ means that $p_3 \sim 0$. In this regard, by the use of Eq.~(\ref{miuprima-1D}) we can approximate the Bose-Einsten distribution $n^{+}(p_3)$ in a vicinity of $p_3 =0$ as

\begin{eqnarray}
n^+(p_3) = \frac{1}{e^{\beta(\varepsilon(p_3,B)-\mu)-1}} \simeq \frac{2 \varepsilon(0,B) T}{p^2 +\varepsilon(0,B)^2 - \mu^2} \simeq \frac{2 \varepsilon(0,B) T}{p^2 -2 \varepsilon(0,B) \mu^{\prime 1D}}, \\\nonumber
\end{eqnarray}

\begin{eqnarray}\label{nmas}
n^+(p_3) \simeq 2 N \frac{\gamma}{p_{3}^2 + \gamma^2},
\end{eqnarray}

\noindent where $\gamma = \sqrt{- 2 \varepsilon(0,B) \mu^{\prime 1D}}$. Eq.(\ref{nmas}) means that for $p_3 \approx 0$, $n^+(p_3)$ tends to a
Cauchy distribution centered in $p_3=0$ (as happens for the charged vector bosons distribution \cite{ROJAS1996148}). Now the equivalent to the limits  $T \rightarrow 0$ or  $B\rightarrow B_c$  is the limit $\gamma \rightarrow 0$  and

\begin{equation}\label{nmas1}
\lim_{\gamma\rightarrow 0} n^+(p_3) = 2 \pi N \delta(p_3).
\end{equation}

Using Eq.(\ref{nmas1}) we have the following expression for the particle density $N$ in the vicinity of $p_3=0$

\begin{equation}\label{densityTcero}
N \simeq \frac{1}{2 \pi} \lim_{\gamma\rightarrow 0}\int_{-\infty}^{\infty} 2 \pi N \frac{\gamma}{p_{3}^2 + \gamma^2} dp_3 = N \int_{-\infty}^{\infty} \delta(p_3)dp_3.
\end{equation}

 The delta behaviour of the Bose-Einstein distribution $ n^+(p_3)$ for low temperatures or high fields is depicted in Fig.~3.
Left panel of Fig.~3 shows the boson distribution as a function of $p_3$ for a fixed value of temperature $T=10^7 K$ and several values of the magnetic field. The maximum of the curves increases and move to the left when the magnetic field grows.
Right panel of Fig.~3 shows the boson distribution as a function of $p_3$ for a fixed value of the magnetic field $B=0.1 B_c$ and  different values of the temperatures.  The maximum of the curves increases and shift to the left with the decrease of the temperature.
Fig.~3 illustrates that although there is not a critical temperature (or field), as the limits $T \rightarrow 0$ or $B \rightarrow B_c$ are approached, the bosons concentrate in the ground state and its neighboring states. Therefore, the system has a diffuse BEC.
	
\begin{figure}[t]
\centering
\includegraphics[width=0.49\linewidth]{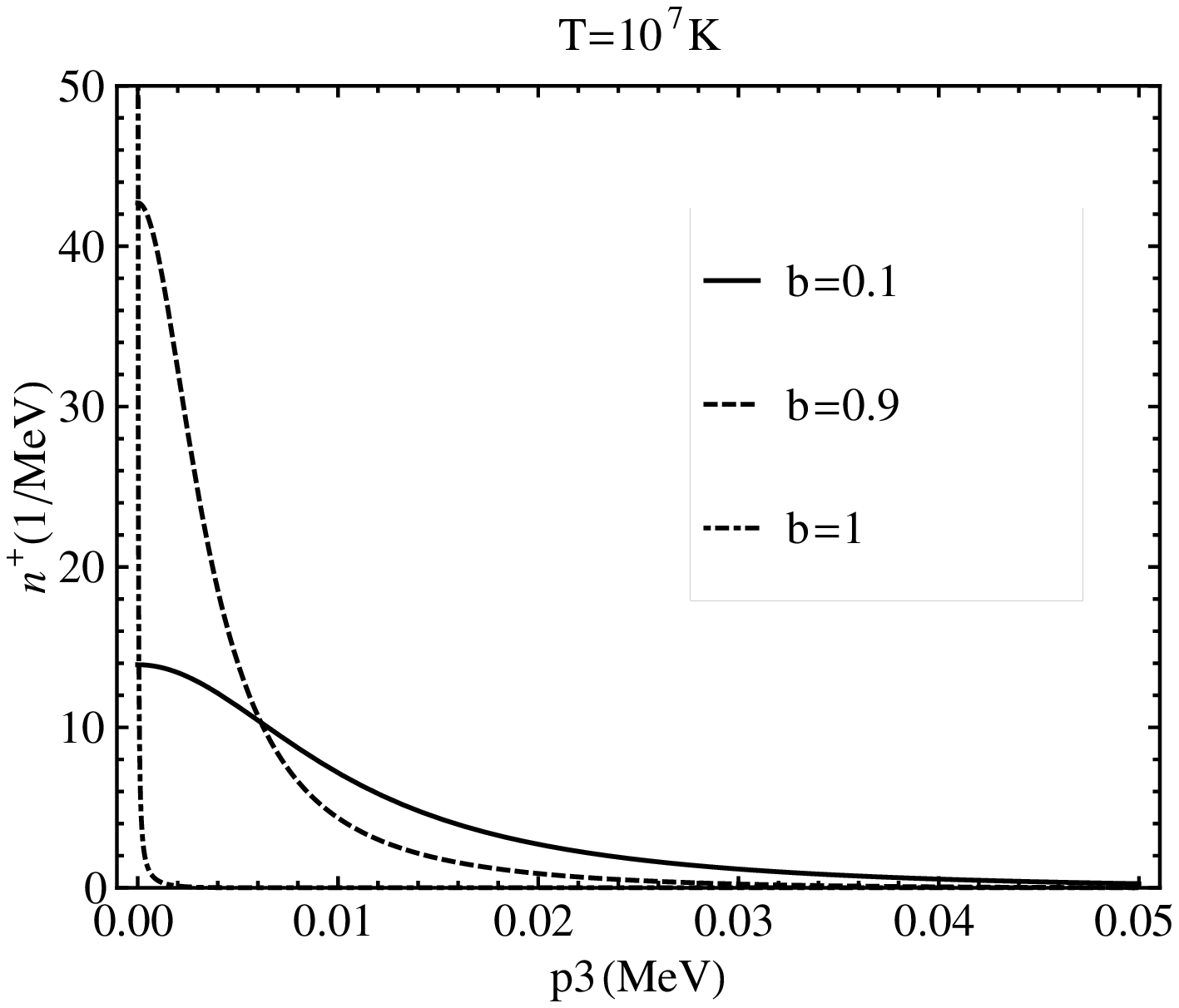}
\includegraphics[width=0.49\linewidth]{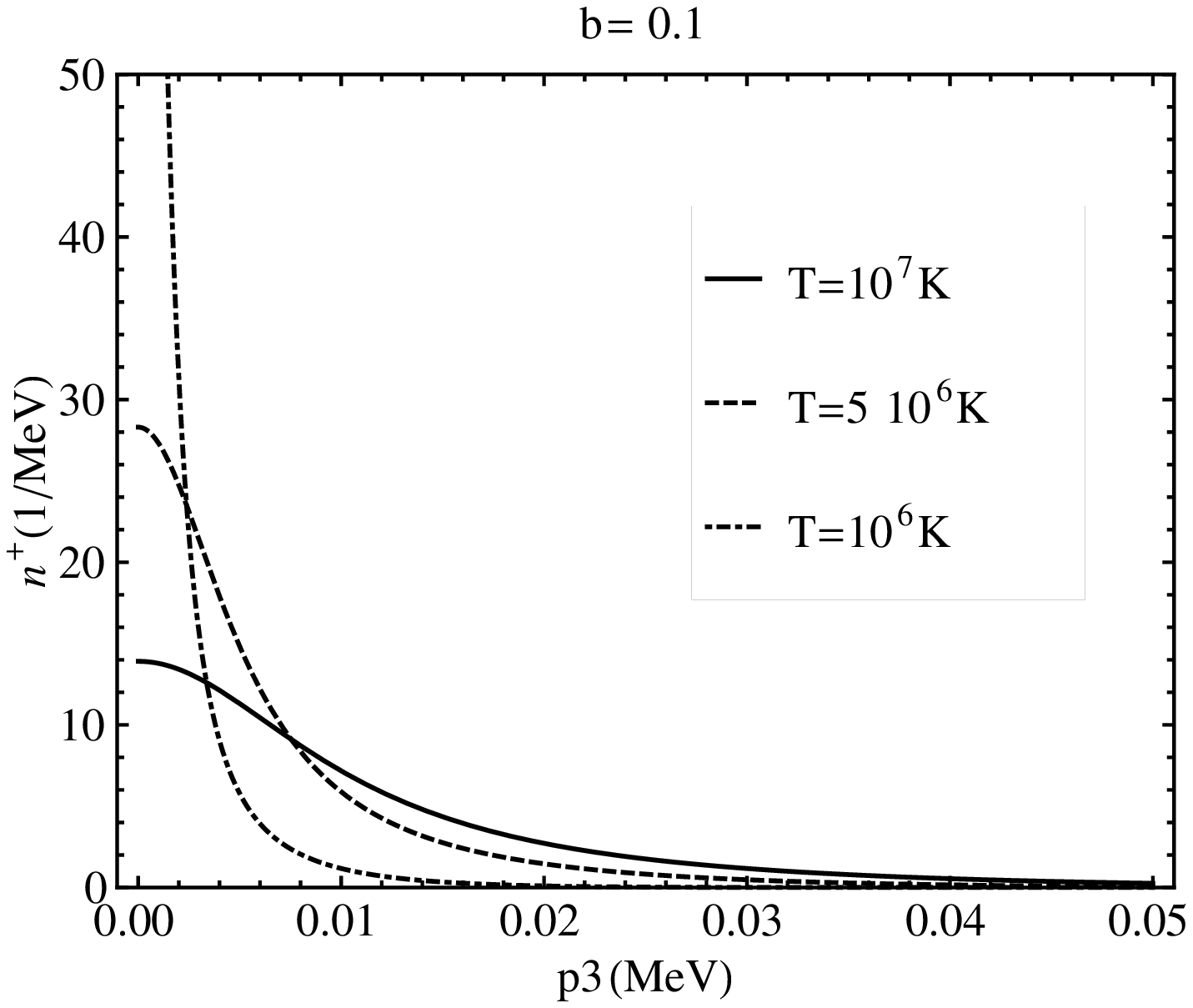}	
\caption{\label{fig1}  Density of bosons as a function of $p_3$ for several values of magnetic field(left plot) and density of boson as a function of $p_3$ for different values of temperatures  (right plot). In both graphics $N=10^{34} cm^{-3}$.}
\end{figure}

The region of temperatures around which the diffuse phase transition occurs may be estimated from the specific heat. To compute the specific heat, we need the energy and the entropy of the one dimensional gas. They read as

\begin{equation}\label{entropy-1D}
S^{1D}=-\frac{\partial \Omega^{1D}}{\partial T}= -\frac{1}{T}\left(\mu^{\prime} N +\frac{3}{2} \Omega^{1D}_{st}+ \frac{2 \varepsilon(0,B)T^2}{N}\right),
\end{equation}

\begin{equation}\label{energy-1D}
E^{1D} = T S^{1D}+\Omega^{1D}+\mu N = \varepsilon(0,B) \left ( N - \frac{2 T^2}{N}\right ) + \Omega^{1D}_{vac}- \frac{1}{2} \Omega^{1D}_{st}.
\end{equation}

From Eq.(\ref{energy-1D}) the specific heat is

\begin{equation}\label{specificheat-1D}
C^{1D}= \frac{\partial E^{1D}}{\partial T} = -\frac{1}{2 T} \left ( \mu^{\prime} N +\frac{3}{2} \Omega_{st} + 10 \frac{\varepsilon(0,B) T^2}{N}\right ).
\end{equation}

\noindent Eq.(\ref{specificheat-1D}) has been plotted as a function of temperature in the right panel of Fig.~4. From this figure it can be seen that the specific heat has a maximum. Is position on the abscise axis can be computed as a function of $b$

\begin{equation}\label{Tc-1D}
T_{max} = \frac{(\zeta (\frac{3}{2}) N)^2}{144 \varepsilon(0,B)}.
\end{equation}

\noindent As $T_{cond}$, $T_{max}$ increases with the density and diverges with the magnetic field. Left panel of Fig.~4 shows $T_{max}$ as a function of $b$ for $N = 10^{34}cm^{-3}$. Although the condensed and not condensed region were shaded, it is worth not to forget that $T_{max}$ does not define a critical temperature. Therefore, left panel of Fig.~4 is only an approximate phase diagram whose provides a range of temperatures around which the population of the ground state starts to grow. The values of these temperatures are in the order of those of several astronomical objects.

\begin{figure}[h]
\centerline{
\includegraphics[width=0.5\linewidth]{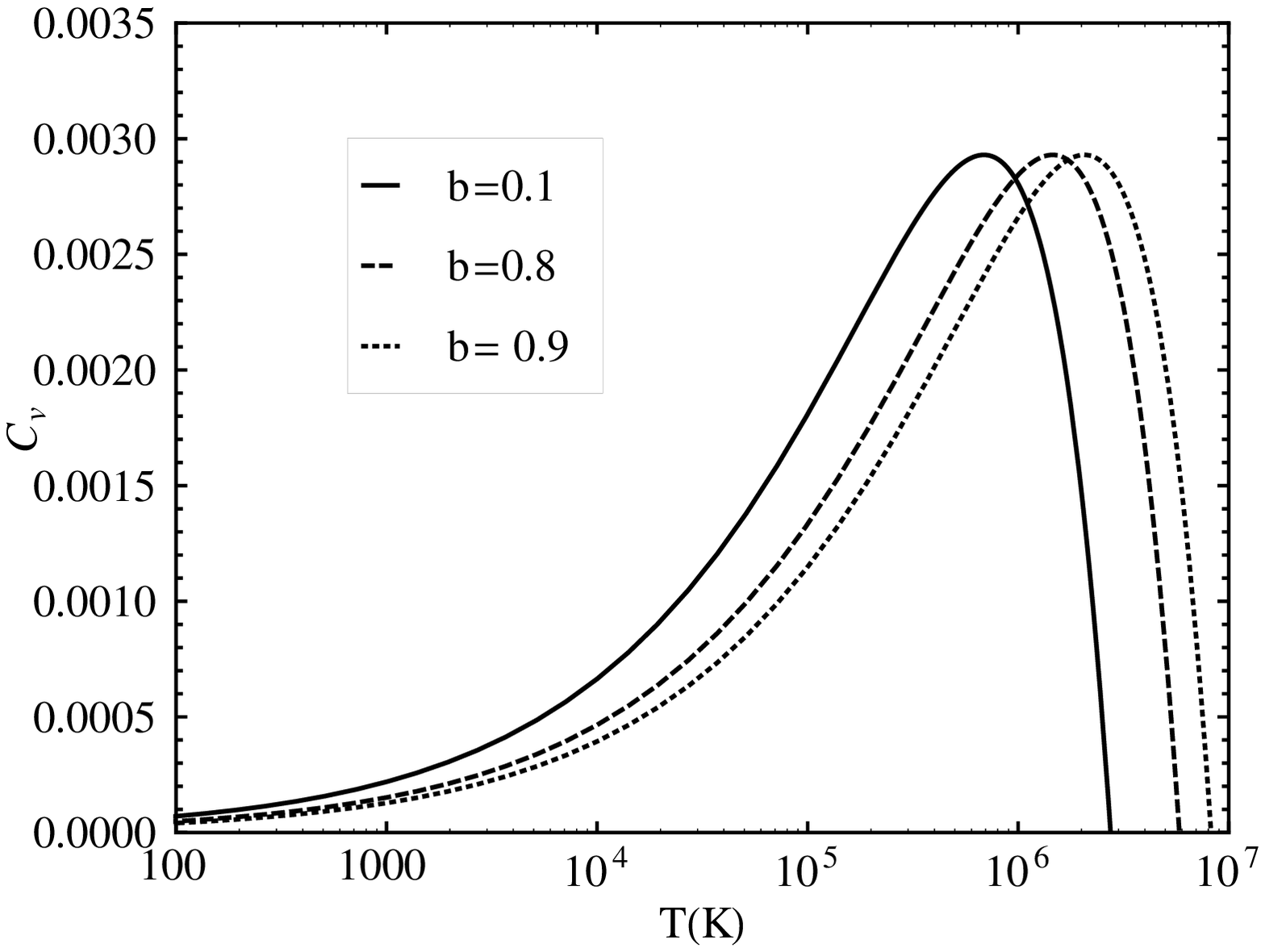}
\includegraphics[width=0.5\linewidth]{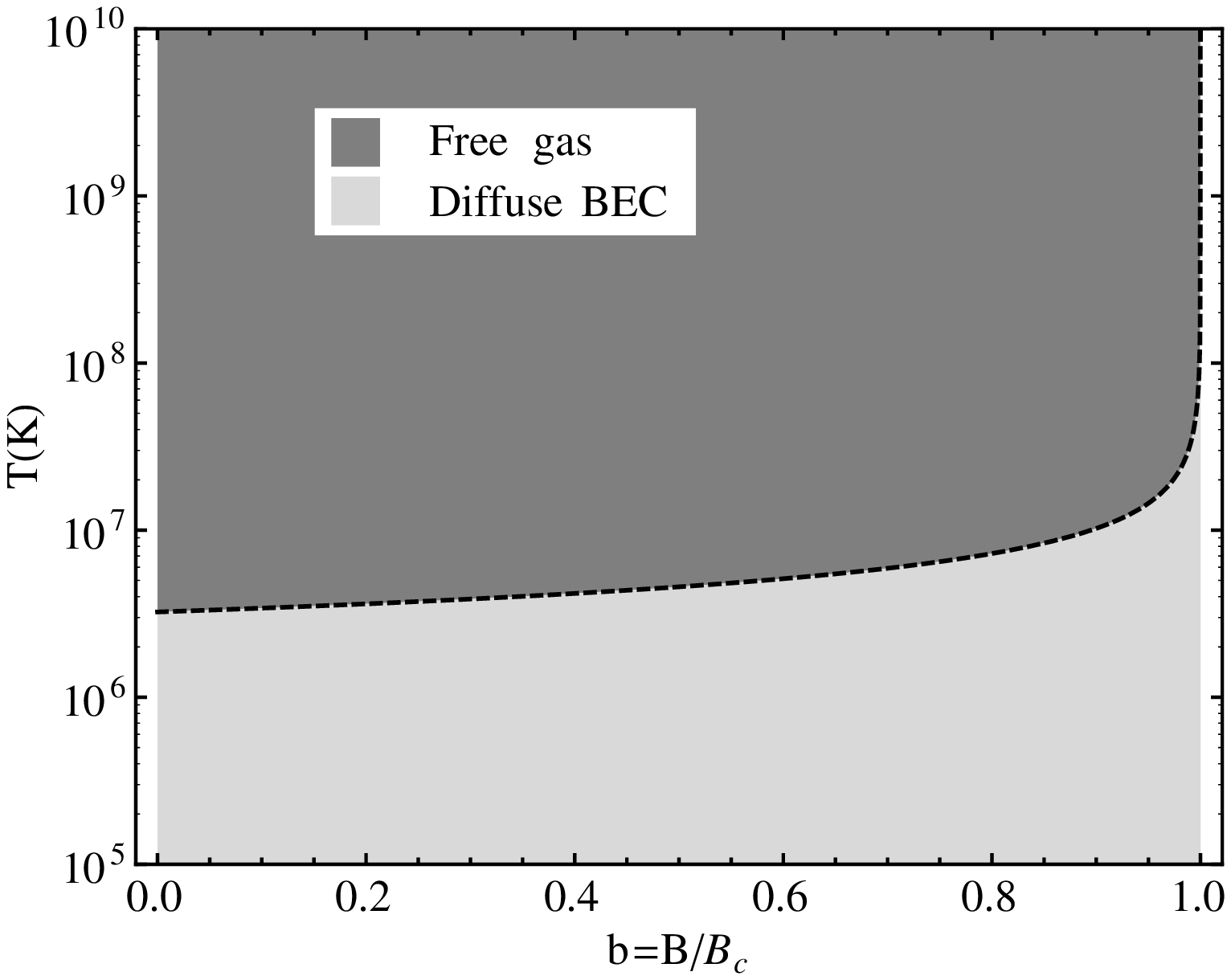}}	
\vspace*{8pt}
\caption{\label{fig4} The specific heat as a function of the temperature (left panel); $T_{max}$ vs. $b$. (right panel). In both graphics $N= 10^{34} cm^{-3}$}
\end{figure}

\subsection{Magnetization}

We can obtain the magnetization of the three and one dimensional systems from Eq.~(\ref{Grand-Potential-Tcero}) and Eq.~(\ref{Grand-Potential-2DstTcero}) if we derive in both expressions with respect to the magnetic field

\begin{equation}\label{magnetization}
M = d N_0 -\frac{\partial \Omega_{st}}{\partial B}-\frac{\partial \Omega_{vac}}{\partial B},
\end{equation}

\begin{equation}\label{magnetization-1D}
M^{1D}=-\frac{\partial \Omega^{1D}_{st}}{\partial B}-\frac{\partial \Omega^{1D}_{vac}}{\partial B}.
\end{equation}

In Eq.~(\ref{magnetization}) $N_0$ is the number of particles in the condensate and $d$ is the effective magnetic moment Eq.(\ref{magmoment}). This term has to be added because in the low temperature limit all the bosons are aligned to the field and contributes to the magnetization, but for temperatures under $T_{cond}$, $\Omega_{st}$ only accounts for the particles that are out of the condensate.

For the statistical contributions to the magnetization we have the expressions

\begin{equation}\label{magnetization-st}
M_{st}=d N_0 -\frac{\partial \Omega_{st}}{\partial B}= \frac{\kappa m}{\varepsilon(0,B)} N - \frac{2 \kappa m T^{5/2}}{(4 \pi)^{5/2} (2-b)^2 \varepsilon(0,B)^{1/2}} Li_{5/2}(e^{\beta \mu^{\prime}}),
\end{equation}

\begin{equation}\label{magnetization-1Dst}
M^{1D}_{st}=-\frac{\partial \Omega^{1D}_{st}}{\partial B}= \frac{\kappa m}{\varepsilon(0,B)} N - \frac{\kappa m T^{3/2} }{2^{3/2} \pi^{1/2} \varepsilon(0,B)^{3/2}} Li_{3/2}(e^{\beta \mu^{\prime}}),
\end{equation}

\noindent while the vacuum contributions are

\begin{equation}\label{magnetization-vac}
M_{vac}= -\frac{\kappa m^3}{72 \pi} \left( 7 b(b^2-6) + 3(2 b^2+2 b-7)(1-b)\log(1-b)-3(2b^2-2b-7)(1+b)\log(1+b)\right ),
\end{equation}

\begin{equation}\label{magnetization-1Dvac}
M^{1D}_{vac}= \frac{\kappa m}{\pi} \log \left(\frac{1+b}{1-b}\right ).
\end{equation}

It is possible to show that for $T\ll m$ the second terms in Eq.(\ref{magnetization-st}) and Eq.(\ref{magnetization-1Dst}) are negligible, as well as one can prove that the vacuum magnetization (Eqs. (\ref{magnetization-vac}) and (\ref{magnetization-1Dvac})) is only relevant for low particle densities at very high fields, so it can be also neglected. Finally, for both, one and three dimensions, the total magnetization of the NVBG is

\begin{equation}\label{magnetizationtotal}
M = \frac{\kappa m}{\varepsilon(0,B)} N =d N.
\end{equation}

The previous expression is expected because at $T\ll m$ all the particles are in the $s=1$ state.  It is nothing else but the product of the effective magnetic moment by the particle density. However, an increase in the field still augment the magnetization because the effective magnetic moment $d$ grows with $B$ and diverges when $B\rightarrow B_c$ ($b \rightarrow 1$). Since $d$ is strictly positive for all values of $B$, the magnetization is always positive and different from zero even if $B=0$ ($M(B=0)=\kappa N$).  This is an evidence of ferromagnetic response of the NVBG at low temperature. This behavior described for $M$ is shown in left panel of Fig.~5.

\begin{figure}[h]
\centering
\includegraphics[width=0.49\linewidth]{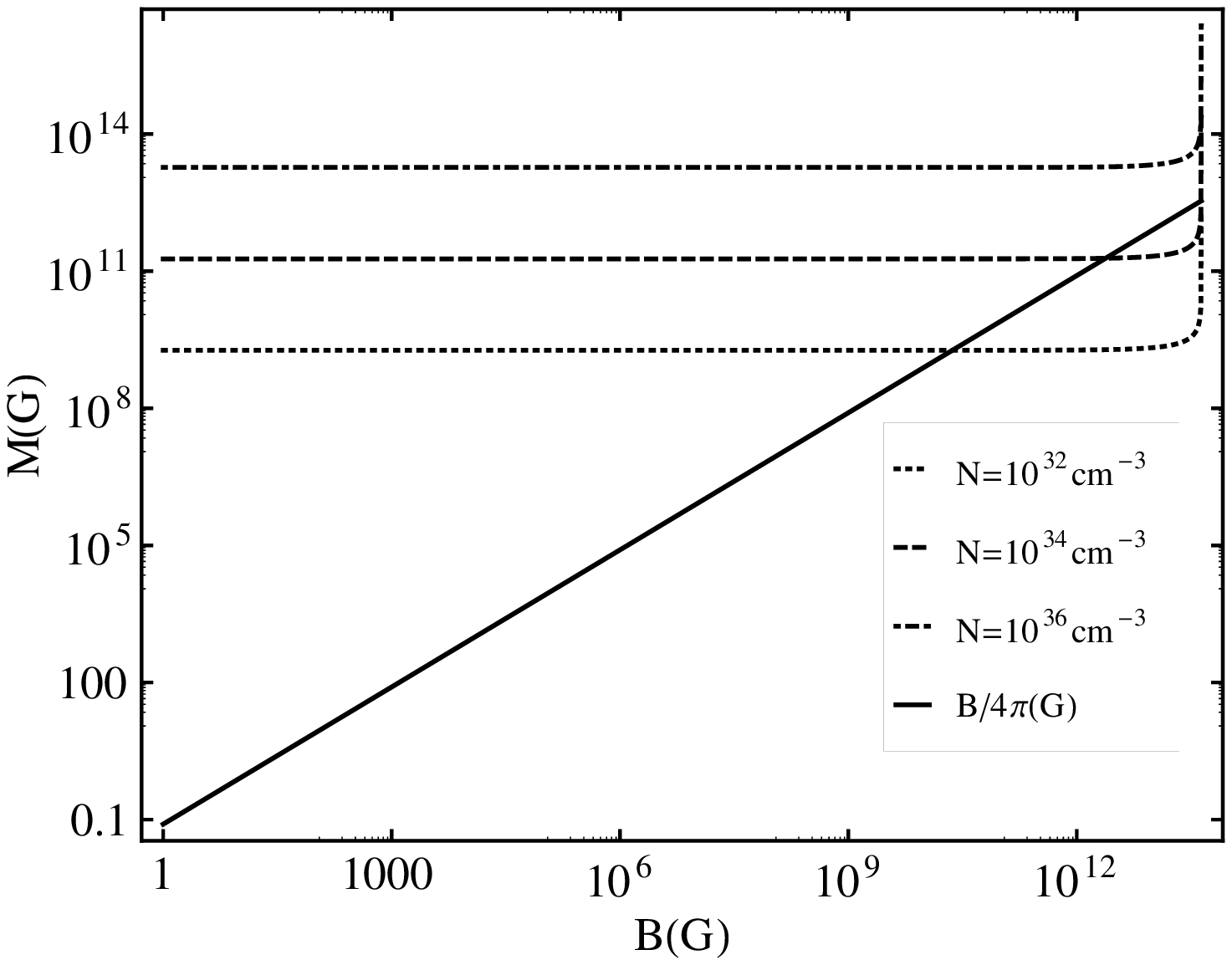}
\includegraphics[width=0.49\linewidth]{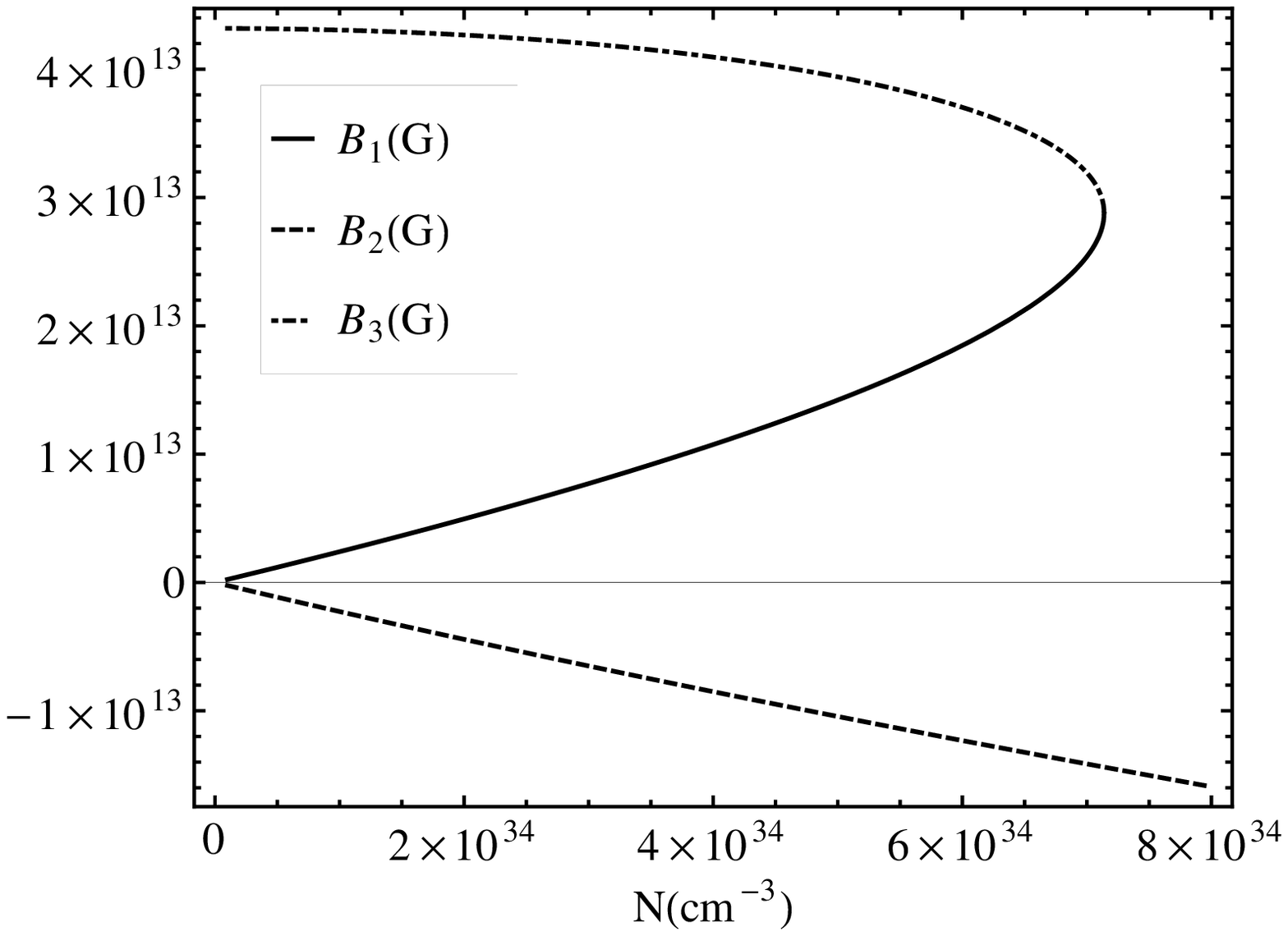}
\caption{\label{fig3} Magnetization as a function of magnetic field for several values of the particle density (left panel). We have also plotted the $B/4\pi$. The solutions of self-magnetization equation as a function of particle density (right panel).}
\end{figure}

 In the seek of one of our main motivations, the search for astrophysical magnetic field sources, we are interested in exploring if the system reaches the self-magnetization condition, i.e. whether or not the solid line in left panel of Fig.~5 intersects the curves of the magnetization. To do that we consider $H = B-4 \pi M$ with no external magnetic field $H=0$, and solve the self-consistent equation $B=4\pi M$. This is a cubic equation due to the non linear dependency of the magnetization on the field. In right panel of Fig.~5 its three solutions have been plotted but only one of them is physically meaningful. For one of the roots, the magnetic field is negative (see dotted line), while for another, it decreases with the increasing density, reaching $B_c$ when $N$ goes to zero (dot dashed line). These solutions implies that the magnetization also decreases with $N$, hence, they are contrary to Eq.~(\ref{magnetizationtotal}) and must be discarded. Therefore, the only admissible solution of the self-magnetization equation is the one given by the solid line. The points of this line are the values of the self-maintained magnetic field. Nevertheless, this solution becomes complex for densities higher than $N_c = 7.14 \times 10^{34}cm^{-3}$. $N_c$ bounds the values of particle densities for which self-magnetization is possible. The maximum field that could be self sustained by the gas corresponds to the critical density and has a magnitude of $2/3\times B_c$. The values of $B$ and $N$ for which a self-magnetization may occur are in the order of those typical of compact objects. The maximum field that can be self maintined by the NVBG is the same obtained for a gas of charged vector bosons with the same mass and magnetic moment, but in this case the critical particle density is of the order of $10^{32}cm^{-3}$ \cite{Elizabeth}.

\subsection{Anisotropic Pressures}

We will consider the energy momentum tensor and the anisotropic pressures of the system.
The total energy momentum tensor of matter plus vacuum will be obtained as a diagonal tensor whose spatial part contains the pressures and the time component is the internal energy density $E$. One gets from the thermodynamical potential
\begin{equation}\label{emtensor}
T^i_j=\frac{\partial\Omega}{\partial a_{i,\lambda}}a_{j,\lambda}-\Omega\delta_{j}^i,\quad\quad T_4^4=-E,
\end{equation}

\noindent where  $a_{i}$ denotes the boson or fermion fields \cite{PerezRojas:2006dq}.
 For a thermodynamical potential that depends on an external field, Eq.(\ref{emtensor}) leads to pressure terms of form

\begin{equation}
\textit{T}^i_j=-\Omega-F_k^i\left( \frac{\partial\Omega}{\partial F_k^j}\right),\quad  i=j.
\end{equation}

Computing the pressures along each direction makes the anisotropy explicit

\begin{eqnarray}\label{pressures}
P_3=\textit{T}_3=-\Omega = -\Omega_{st} -\Omega_{vac},\\
P_{\perp}=\textit{T}_1^1=\textit{T}_2^2=\textit{T}_{\perp}=-\Omega-BM = P_3-BM. \nonumber
\end{eqnarray}

%

In left panel of Fig.~6 the perpendicular and parallel pressures are depicted as function of the field (Eqs. (\ref{pressures})) for $T=10^9 K $ and $N=10^{33} cm^{-3}$. We also shows the statistical and the vacuum parts of the parallel pressure in dashed and dot-dashed lines respectively. The values of the parallel pressure and its statistical part ($-\Omega_{st}$) coincides for $B=0$, but their behavior is different when the field grows. Both are always positive but the total parallel pressure increases with the field and tends to the vacuum contribution $-\Omega_{vac}$, while its statistical part decreases and goes to zero for $B=B_c$ -when all the particles are condensed the gas exerts no pressure. We would like to remark that the parallel pressure remains different form zero due to the vacuum contribution.

On the contrary, the perpendicular pressure decreases (dashed line in left panel of Fig.~6) whit the magnetic field and eventually reaches negative values. This is because the main contribution to $P_{\perp}$ comes from the term $-M B$ which is always negative and diverges in the critical field. A similar result is obtained  for fermion gases in a magnetic field \cite{PerezRojas:2006dq}-\cite{Ferrer:2015wca}. In this frame, a negative pressure can be interpreted as the system becoming unstable. Because the effect of the negative perpendicular pressure is to push the particles inward to the magnetic field axis, we could be in presence of a transversal magnetic collapse \cite{Chaichian}.

Whether the transversal pressure is negative or not depends on the field but also on the temperature and the particle density. This can be seeing if we examine this pressure in more detail for the self-magnetized NVBG. To do that  we substitute in $P_{\perp}$  the solution of the self-magnetization condition $B = 4 \pi M$, and plot the perpendicular pressure as a function of the particle density for several values of $T$ (rihgt panel of Fig.~6). If we start from the lower values of $N$ adding particles to the system increments the parallel pressure. But it also increases the self produced magnetic field and $T_{cond}$. Once $T_{cond}$ becomes higher than the gas temperature, the BEC phase appears and the pressure diminishes because a fraction of the particles fall in the condensate. Besides, as the self-generated field becomes higher, the  contribution to $P_{\perp}$ of the negative term $-M B$ becomes more and more relevant until eventually adding more particles makes the system unstable. A decrease in the temperature lowers the value of particle density where the instability starts.

When the gas is not self-magnetized, but subject to an external magnetic field, an increment in the density continuously also leads the system to the instability. In this case, the increase of $N$ does not augment the field, but it still increments the magnetization and the condensation temperature. Therefore, the NVBG will be unstable or not depending on the values of the temperature, the density and the magnetic field, regardless this field is self produced or not.

The arising of an instability in the magnetized NVBG might be relevant in the description of some phenomena, as jets, that are related to the exertion of mass and radiation out of astronomical objets \cite{Elizabeth}.

\begin{figure}[h!
 ]
\centering
\includegraphics[width=0.49\linewidth]{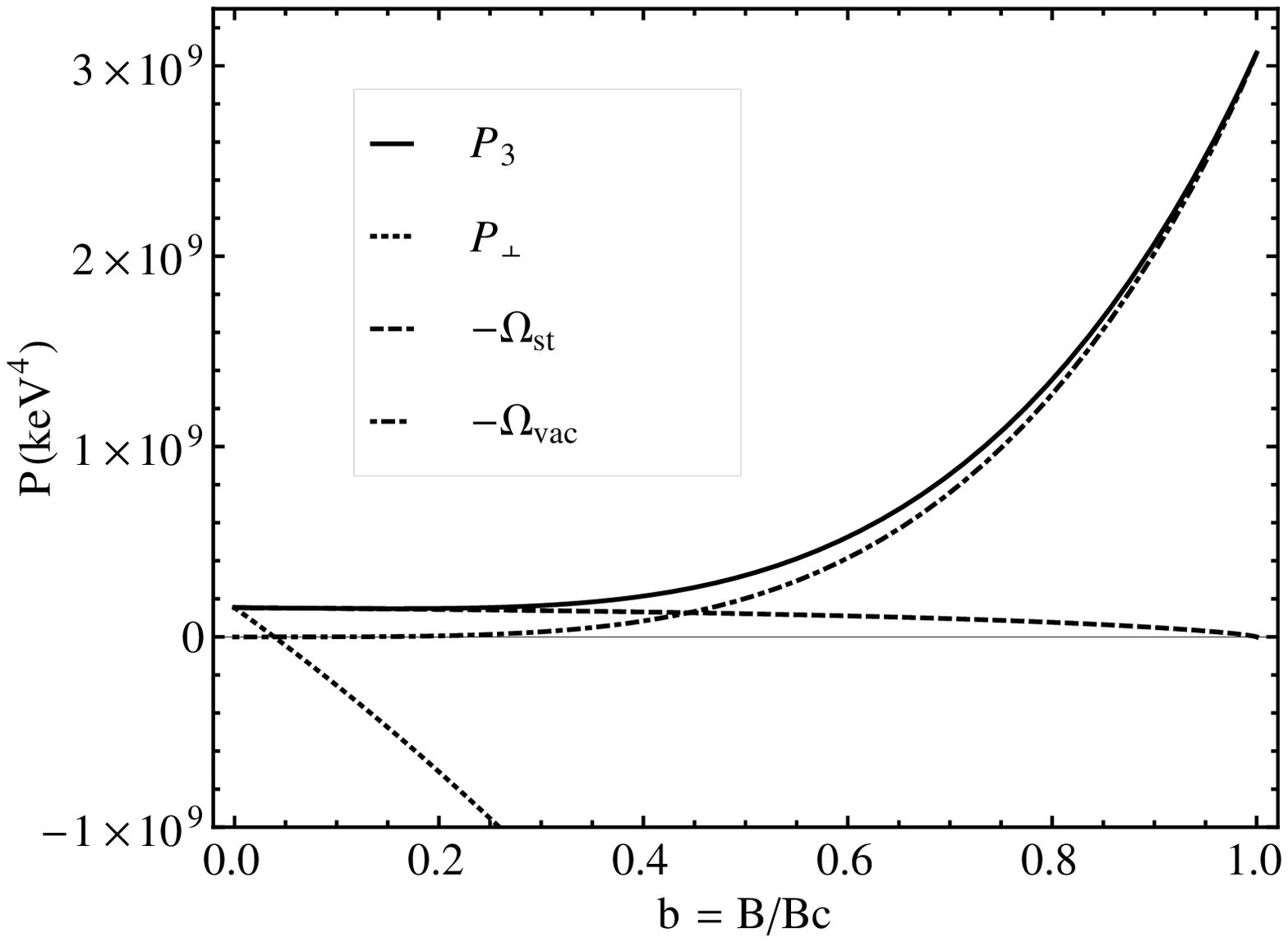}
\includegraphics[width=0.49\linewidth]{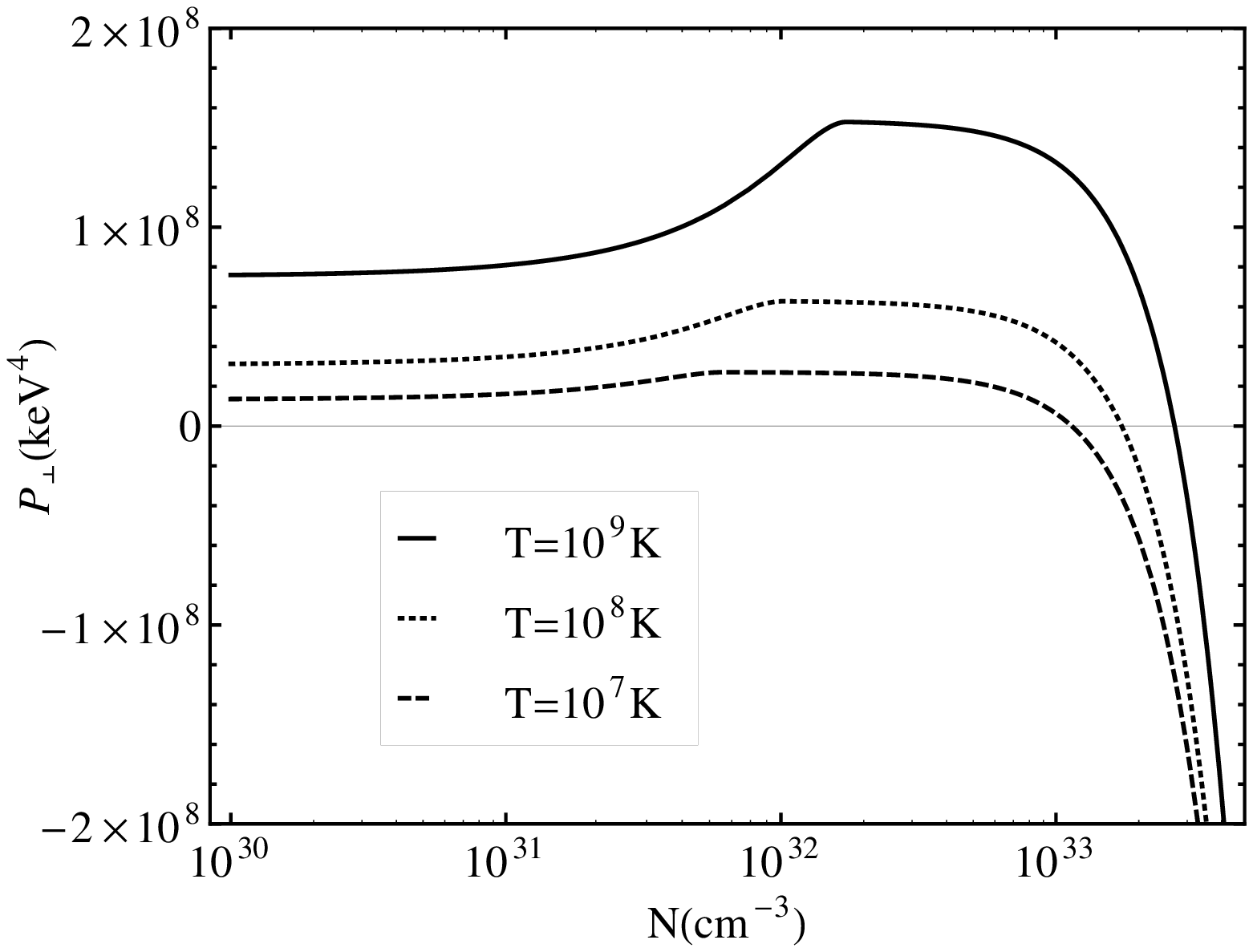}	
\caption{\label{fig8} The pressure as a function of the magnetic field for several values of temperature (left panel); the statistical and the vacuum contributions to the pressure are also plotted in dashed and dot-dashed lines. The perpendicular pressure of the self-magnetized gas as a function of the particle density for several values of temperature (right panel).}
\end{figure}

\section{Conclusions}\label{conclusions}
Starting from the Proca formalism  \cite{PhysRev.131.2326,PhysRevD.89.121701} we computed the spectrum of a gas of neutral vector bosons in a constant magnetic field. The effective rest energy Eq.~(\ref{massrest}) turns out to be a decreasing function of the magnetic field that becomes zero when it reaches certain critical value $B_c=\frac{m}{2 k}$.

When the temperature is low enough, the NVBG undergoes a phase transition to a Bose-Einstein Condensation. In dependence on whether the gas is three or one dimensional, this transition is usual or diffuse. However, in one dimension as well as in three, the phase transition to the BEC is driven not only by the temperature or the density, but also by the magnetic field.

The magnetization of the gas is a positive quantity that increases with the field and diverges when $B=B_c$ for both, the three and the one dimensional cases.
For particle densities under a critical value $N_c \cong 7.14 \times 10^{34}cm^{-3}$ the self-magnetization condition is fulfilled and the gas can maintain a self-generated magnetic field. The maximum field that can be reached by self-magnetization turns out to be $2/3 \times B_c \sim 10^{13} G$.

The change of spherical to axial symmetry induced by the magnetic field is explicitly manifested in the spectrum of the NVBG (through the asymmetry in the momentum components) and also in the splitting of the pressures in the parallel and perpendicular directions to the field. For low values of the field, the pressure exerted by the particles has the main role in both components. However, when the magnetic field grows, the increasing parallel pressure is dominated by the positive vacuum pressure term, while the decreasing perpendicular pressure is determined by the negative magnetic pressure term $-M B$. For magnetic fields and particle densities high enough, or low enough temperatures, the perpendicular pressure becomes negative and an instability emerges in the system that turns out to be susceptible to suffer a transversal magnetic collapse.

All these phenomena undergone by NVBG (BEC, self-sustained magnetic field and the collapsing  of the transverse pressure) appear for typical values of densities and magnetic fields in compact objects. Therefore they could be relevant in modeling jets as well as the mechanism that  sustain the strong magnetic field in compact objects. These models deserve a separated treatment which is in progress.


\section{Acknowledgements}

The authors thank the comments of Maxim Chernodub to the first version of this work.
G.Q.A, A.P.M and H.P.R have been supported by the grant CB0407 and acknowledge the receipt of the grant from the Abdus Salam International Centre for Theoretical Physics, Trieste, Italy.

\appendix

\section{Calculation of $I$}

In order to compute the integral of the second term of Eq.(\ref{Grand-Potential-sst2})

\begin{equation}\label{I3}
I=\int\limits_{z_0}^{\infty} dz \frac{z^2}{\sqrt{z^2+\alpha^2}} K_1 (y z),
\end{equation}

\noindent let's introduce the following form for $K_1 (yz)$

\begin{equation}\label{k1int}
K_1 (yz) = \frac{1}{y z}\int\limits_{0}^{\infty} dt e^{-t-\frac{y^2 z^2}{4 t}}.
\end{equation}

If we substitute (\ref{k1int}) in (\ref{I3}), the integration over $z$ can be carried out

\begin{equation}\label{I31}
I= \frac{\sqrt{\pi}}{y^2} \int\limits_{0}^{\infty} dt \sqrt{t} e^{-t+\frac{y^2 \alpha^2}{4 t}} erfc \left(\frac{y \sqrt{z^2+\alpha^2}}{2 \sqrt{t}}\right).
\end{equation}

To integrate over $t$ in (\ref{I31}) we replace the complementary error function $erfc(x)$ by its series expansion

\begin{equation}\label{erfc}
erfc(x) \backsimeq \frac{e^{-x^2}}{\sqrt{\pi} x} \left(1 - \sum_{w=1}^{\infty} \frac{(-1)^w(2 w -1)!!}{(2 x^2)^w}\right).
\end{equation}

After the replacement and integration, $I_3$ can be written as

\begin{equation}\label{I32}
I = \frac{z_0^2}{y \sqrt{z_{0}^2 + \alpha^2}} K_2 (y z_0) -
\frac{z_0^2}{y \sqrt{z_{0}^2 + \alpha^2}} \sum_{w=1}^{\infty} \frac{(-1)^w(2 w -1)!!}{(z_0^2+\alpha^2)^w} \left(\frac{z_0}{y}\right)^w K_{-(w+2)} (y z_0).
\end{equation}

\section{Vacuum thermodynamical potential}

To obtain Eq.(\ref{Grand-Potential-vac}) for the vacuum contribution to the thermodynamical potential we start from its definition

\begin{equation}\label{Grand-potential-vac-1}
\Omega_{vac}=\sum_{s=-1,0,1}\int\limits_{0}^{\infty}\frac{p_{\perp}dp_{\perp}dp_3}{(2\pi)^2}\varepsilon \nonumber
\end{equation}

where $\varepsilon(p_{\perp},p_3, B,s)=\sqrt{p_3^2+p_{\perp}^2+m^2-2\kappa s B\sqrt{p_{\perp}^2+m^2}}$.

To integrate over $p_3$ and $p_{\perp}$ we use the equivalence

\begin{equation}
\sqrt{a}= -\frac{1}{2 \sqrt{\pi}} \int\limits_{0}^{\infty} dy y^{-3/2} (e^{- y a}-1)
\end{equation}

\noindent and introduce the small quantity $\delta$ as lower limit of the integral to regularize the divergence of the $a$ dependent term and eliminate the term that does not depends on $a$

\begin{equation}
\sqrt{a(\delta)}= -\frac{1}{2 \sqrt{\pi}} \int\limits_{\delta}^{\infty} dy y^{-3/2} e^{-y a}.
\end{equation}

Now, let's make $a(\delta) = \varepsilon^2 = p_3^2+p_{\perp}^2+m^2-2\kappa s B\sqrt{p_{\perp}^2+m^2}$. Consequently

\begin{equation}\label{energyintegral}
\varepsilon = -\frac{1}{2 \sqrt{\pi}} \int\limits_{\delta}^{\infty} dy y^{-3/2} e^{- y(p_3^2+p_{\perp}^2+m^2-2\kappa s B\sqrt{p_{\perp}^2+m^2})}.
\end{equation}

By substituting Eq.(\ref{energyintegral}) in Eq.(\ref{Grand-potential-vac-1}) we obtain for the vacuum thermodynamical potential

\begin{equation}\label{Grand-potential-vac-2}
\Omega_{vac}=-\frac{1}{8 \pi^{5/2}}\sum_{s=-1,0,1}\int\limits_{\delta}^{\infty} dy y^{-3/2} \int\limits_{0}^{\infty}dp_{\perp} p_{\perp} \int\limits_{-\infty}^{\infty}dp_3 e^{- y(p_3^2+p_{\perp}^2+m^2-2\kappa s B\sqrt{p_{\perp}^2+m^2})}
\end{equation}

After doing the gaussian integral over $p_3$ Eq.(\ref{Grand-potential-vac-2}) reads

\begin{equation}\label{Grand-potential-vac-3}
\Omega_{vac}=-\frac{1}{8 \pi^{2}}\sum_{s=-1,0,1}\int\limits_{\delta}^{\infty} dy y^{-2} \int\limits_{0}^{\infty}dp_{\perp} p_{\perp} e^{- y(p_{\perp}^2+m^2-2\kappa s B\sqrt{p_{\perp}^2+m^2})}.
\end{equation}

If we introduce the new variable $z = \sqrt{m^2+p_{\perp}^2} - s \kappa B$, Eq.(\ref{Grand-potential-vac-3}) becomes

\begin{equation}\label{Grand-potential-vac-4}
\Omega_{vac}=-\frac{1}{8 \pi^{2}}\sum_{s=-1,0,1} \left \{\int\limits_{\delta}^{\infty} dy y^{-3} e^{- y(m^2-2 m s \kappa B)}
+ s \kappa B \int\limits_{\delta}^{\infty} dy y^{-2} \int\limits_{z_1}^{\infty}dz e^{- y(z^2 - s^2 \kappa^2 B^2)}
 \right \},
\end{equation}

\noindent where $z_1 = m - s \kappa B $.

Eq.(\ref{Grand-potential-vac-4}) admits a further simplification if we perform a second change of variables $w=z-z_1$ in its last term, sum over the spin and remember that $b = B/B_c$ with $B_c = m / 2 \kappa$

\begin{equation}\label{Grand-potential-vac-5}
\Omega_{vac}=-\frac{1}{8 \pi^{2}}\left \{ \int\limits_{\delta}^{\infty} dy y^{-3} e^{- ym^2} (1+2 \cosh{[m^2 b y]})
+ m b \int\limits_{\delta}^{\infty} dy y^{-2} \int\limits_{0}^{\infty}dw e^{- y(m - w)^2} \sinh[m b (m - w) y] \right \}.
\end{equation}

To take the limit $\delta \rightarrow 0$ we subtract from $1+2 \cosh{[m^2 b y]}$ and $ \sinh[m b (m - w) y]$ the first terms in their series expansion and obtain for the vacuum thermodynamical potential the expression

\begin{eqnarray}\label{Grand-potential-vac-6}
\Omega_{vac}=-\frac{1}{8 \pi^{2}}\int\limits_{0}^{\infty} dy y^{-3} e^{- ym^2} \{2 \cosh{[m^2 b y]} - 2 - m^4 b^2 y^2 \}- &\\
-\frac{m b}{8 \pi^{2}} \int\limits_{0}^{\infty} dy y^{-2} \int\limits_{0}^{\infty}dw e^{- y(m - w)^2}\left \{ \sinh[m b (m - w) y]- m b (m - w) y - \frac{[m b (m - w) y]^3}{6} \right \} \nonumber
\end{eqnarray}

\noindent that leads to Eq.(\ref{Grand-Potential-vac}) after integration.

A similar procedure was used to obtain the vacuum contribution to the thermodynamical potential for the one dimensional gas (Eq.~ (\ref{Grand-Potential-2Dvacreg})).


\begin{thebibliography}{1}


\bibitem{Lattimer:2004pg}
  J.~M.~Lattimer and M.~Prakash,
  Science {\bf 304} (2004) 536
  doi:10.1126/science.1090720
  [astro-ph/0405262].
\bibitem{0954-3899-36-7-075202}
R.~G. Felipe and A.~P. Mart\'inez, {\em Journal of Physics G: Nuclear and
  Particle Physics} {\bf 36}, p. 075202  (2009).

\bibitem{MNL2:MNL2848}
J.~Charbonneau, K.~Hoffman and J.~Heyl, {\em Monthly Notices of the Royal
  Astronomical Society: Letters} {\bf 404}, L119  (2010).

\bibitem{Charbonneau:2009ax}
J.~Charbonneau and A.~Zhitnitsky, {\em JCAP} {\bf 1008}, p. 010  (2010).
\bibitem{Yamada} Keiji Yamada, Prog Theor Phys (1982) 67 (2): 443-453. DOI:https://doi.org/10.1143/PTP.67.443
\bibitem{Elizabeth} H. Perez Rojas, E. Rodriguez Querts, A. Perez Martinez  Conference Series (Quantum Relativistic electron gas expanding in one dimension), (2017).


\bibitem{ROJAS1996148}
H.~Rojas, {\em Physics Letters B} {\bf 379}, 148   (1996).

\bibitem{PEREZROJAS2000}
H.~Perez~Rojas and L.~Villegas-Lelovski, {\em Brazilian Journal of Physics}
  {\bf 30}, 410  (06 2000).



\bibitem{Khalilov:1999xd}
  V.~R.~Khalilov and C.~L.~Ho,
  Phys.\ Rev.\ D {\bf 60} (1999) 033003
  doi:10.1103/PhysRevD.60.033003
  [hep-th/0001120].

 \bibitem{Khalilov1997}
  V.~R. Khalilov, C.~L.~Ho and C.~Yang {\em Modern Physics Letters A} {\bf 12}, 1973
  (1997).

%

\bibitem{Bargueno}R. L.  Delgado, R. L, P. Bargue\~no, F.  Sols.	Physical Review E, vol. 86, Issue 3, 10.1103/PhysRevE.86.031102 (2012).
\bibitem{Jian} X. Jian, J. Q.Gu {\em J. Phys. Condens. Matter}, {\bf 23}, 026003 (2011).

\bibitem{ROJAS1997}
H.~Rojas, {\em Physics Letters A} {\bf 234}, 13   (1997).

\bibitem{Chernodub:2010qx}
  M.~N.~Chernodub,
  Phys.\ Rev.\ D {\bf 82} (2010) 085011
  doi:10.1103/PhysRevD.82.085011
  [arXiv:1008.1055 [hep-ph]].

\bibitem{Chernodub:2012fi}
  M.~N.~Chernodub, J.~Van Doorsselaere and H.~Verschelde,
  Phys.\ Rev.\ D {\bf 88} (2013) 065006
  doi:10.1103/PhysRevD.88.065006
  [arXiv:1203.5963 [hep-ph]].


\bibitem{Satarov:2017jtu}
  L.~M.~Satarov, M.~I.~Gorenstein, A.~Motornenko, V.~Vovchenko, I.~N.~Mishustin and H.~Stoecker,
  arXiv:1704.08039 [nucl-th].



\bibitem{PhysRev.131.2326}
J.~A. Young and S.~A. Bludman, {\em Phys. Rev.} {\bf 131}, 2326 (Sep 1963).

\bibitem{PhysRevD.89.121701}
A.~J. Silenko, {\em Phys. Rev. D} {\bf 89}, p. 121701 (Jun 2014).


\bibitem{PerezRojas:2006dq}
  H.~Perez Rojas and E.~Rodriguez Querts,
  Int.\ J.\ Mod.\ Phys.\ A {\bf 21} (2006) 3761
  doi:10.1142/S0217751X06031715
  [hep-ph/0603254].


 \bibitem{Chaichian} M. Chaichian, S. S. Masood, C. Montonen, A. Perez Martinez and H. Perez Rojas, Phys. Rev. Lett. 84 (2000) 5261;A. Perez-Martinez, H. Perez-Rojas, and H. J. Mosquera-Cuesta, Eur. Phys. J. C29, 111 (2003); S. Chakrabarty, \textit{Phys. Rev. D} \textbf{54} (1996) 1306; R. Gonzalez-Felipe, A. Perez Martinez, H. Perez Rojas, and M. Orsaria, Phys. Rev. C 77, 015807 (2008);


\bibitem{Ferrer:2015wca}
  E.~J.~Ferrer, V.~de la Incera, D.~Manreza Paret, A.~P\'erez Mart\'inez and A.~Sanchez,
  Phys.\ Rev.\ D {\bf 91} (2015) no.8,  085041
  doi:10.1103/PhysRevD.91.085041
  [arXiv:1501.06616 [hep-ph]].

\end{thebibliography}

\end{document}